\documentclass[10pt,letterpaper]{article}
\usepackage[top=0.85in,left=1.0in,footskip=0.75in]{geometry} 

\usepackage{rotating}
\usepackage{multirow}
\usepackage{booktabs}
\usepackage{amsthm}
\usepackage{soul}
\usepackage{ulem}
\usepackage[e]{esvect}
\usepackage{float}

\usepackage{amsmath,amssymb}
\usepackage{changepage}

\usepackage[utf8x]{inputenc}

\usepackage{textcomp,marvosym}

\usepackage{cite}

\usepackage{nameref}
\usepackage[hidelinks=true]{hyperref}

\usepackage{microtype}
\DisableLigatures[f]{encoding = *, family = * }

\usepackage[table]{xcolor}

\usepackage{array}

\newcolumntype{+}{!{\vrule width 2pt}}

\newlength\savedwidth

\raggedright
\usepackage[aboveskip=1pt,labelfont=bf,labelsep=period,justification=raggedright,singlelinecheck=off]{caption}

\makeatletter
\renewcommand{\@biblabel}[1]{\quad#1.}
\makeatother

\date{}

\usepackage{lastpage,fancyhdr,graphicx}
\usepackage{epstopdf}

\newcommand{\beginsupplement}{%
        \setcounter{table}{0}
        \renewcommand{\thetable}{S\arabic{table}}%
        \setcounter{figure}{0}
        \renewcommand{\thefigure}{S\arabic{figure}}%
     }

\newcommand{\mcall}[1]{{\fontfamily{qcr}\selectfont
#1}}

\definecolor{pyGold}{RGB}{250,250,210}
\definecolor{DSVL}{RGB}{0,158,115}
\definecolor{SLVL}{RGB}{0,11,178}
\definecolor{SHVL}{RGB}{213,94,0}
\definecolor{HVLS}{RGB}{204,121,167}
\definecolor{RVL}{RGB}{0,0,0}

\usepackage{mathtools}

\begin{document}
\vspace*{0.2in}

\begin{flushleft}
{\Large
\textbf\newline{Revealing patterns in HIV viral load data and classifying patients via a novel machine learning cluster summarization method} 
%
}
\newline
\\

Samir Farooq\textsuperscript{1,2},
Samuel J. Weisenthal\textsuperscript{1,4},
Melissa Trayhan\textsuperscript{1,2,4},
Robert J. White\textsuperscript{1,2,4},
Kristen Bush\textsuperscript{1,4},
Peter R. Mariuz\textsuperscript{3},
Martin S. Zand\textsuperscript{1,2,4,*}
\\
\bigskip
\textbf{1} Rochester Center for Health Informatics, 265 Crittenden Boulevard, Rochester, NY 14642-0708, USA
\\
\textbf{2} Department of Medicine, Division of Nephrology, 601 Elmwood Ave - Box 675, Rochester NY 14642, USA
\\
\textbf{3} Department of Medicine, Division of Infectious Diseases, University of Rochester Medical Center/Strong Memorial Hospital AIDS Center, 601 Elmwood Ave, Rochester NY 14642, USA
\\
\textbf{4} Clinical and Translational Science Institute,
University of Rochester Medical Center, 265 Crittenden Boulevard, Rochester, NY 14642-0708, USA\\
\bigskip

%
%





* martin\_zand@urmc.rochester.edu

\end{flushleft}
\section*{Abstract}
HIV RNA viral load (VL) is an important outcome variable in studies of HIV infected persons. There exists only a handful of methods which classify patients by viral load patterns.  Most methods place limits on the use of viral load measurements, are often specific to a particular study design, and do not account for complex, temporal variation. To address this issue, we propose a set of four unambiguous computable characteristics (features) of time-varying HIV viral load patterns, along with a novel centroid-based classification algorithm, which we use to classify a population of 1,576 HIV positive clinic patients into one of five different viral load patterns (clusters) often found in the literature: durably suppressed viral load (DSVL), sustained low viral load (SLVL), sustained high viral load (SHVL), high viral load suppression (HVLS), and rebounding viral load (RVL). The centroid algorithm summarizes these clusters in terms of their centroids and radii. We show that this allows new viral load patterns to be assigned pattern membership based on the distance from the centroid relative to its radius, which we term radial normalization classification. This method has the benefit of providing an objective and quantitative method to assign viral load pattern membership with a concise and interpretable model that aids clinical decision making. This method also facilitates meta-analyses by providing computably distinct HIV categories. Finally we propose that this novel centroid algorithm could also be useful in the areas of cluster comparison for outcomes research and data reduction in machine learning.


\section*{Introduction}

The primary clinical goal of HIV treatment and patient engagement is suppression of the HIV viral load (VL), as measured by low or undetectable circulating HIV RNA levels. However, VL most often fluctuates over repeated measurements, with a range that spans 8 orders of magnitude from 0 (undetectable) - $10^7$ copies/mL. VL is regularly monitored for signs of progression of HIV infection.  Standard HIV treatment protocols are based on VL measurements \cite{centers2011vital}, especially when monitoring responses to antiretroviral therapy (ART). Monitoring of VL helps to determine whether ART therapy was able to successfully suppress patient viral load \cite{yehia2012sustained}. Individuals with sustained high viral loads (SHVL) are at greater risk of secondary transmission, clinical progression to AIDS, and death\cite{I1,I2,I3,VL400_1}. In contrast, significant reduction in viral load, or high viral load suppression (HVLS), lead to immune recovery, as measured by CD4 T cell levels\cite{SupBenefits1}, and can reduce or eliminate the risks of SHVL. Furthermore, patients sustaining low-level viral load (SLVL), or with a rising viral load after previous suppression, have a high incidence of treatment failure \cite{greub2002intermittent}. Thus, developing an objective measure of  viral load status, and categorization of patients by time varying patterns of viral load, is critical for standardizing both therapy and comparing research protocol efficacy.

Reports in the current literature differ in the definition ``high viral load" \cite{main,rose2015comparison,de2006fatal,ylitalo2000consistent}, and their findings of how long it takes a patient on HAART to suppress their viral load \cite{main,yehia2012sustained,rose2015comparison,phillips2001hiv}.  We summarize some of the published approaches here (for greater detail see Supplementary Material).  With respect to viral load levels, Terzian et. al. defined SHVL as two consecutive viral load measurements (VLM) $\geq$100,000 copies/mL \cite{main}. Furthermore, they define durably suppressed viral load (DSVL) as all VLM $<$400 copies/mL. In contrast, Greub et. al.  focused on detecting low level viral rebound (LLVR) in patients by first considering patients with an initial consecutive VLM pair $<$50 copies/mL, and classified LLVR as having subsequent maximum VLM between 51-500 \cite{greub2002intermittent}. Alternatively, Rose et. al. investigated the use of five different frameworks to categorize suppressed versus not-suppressed viral load \cite{rose2015comparison}. Their approach excluded patients with VLM$<$200 at baseline, and stratified the remainder with regard to VL suppression using an 8 month window centered around 24 months after the start of VLM (18-30 months).  Phillips et al. characterized VLM responses to ART \cite{phillips2001hiv}, utilizing a 24-40 week window and a rule-based method to identify two populations of HIV patients (Viral Failure and Viral Rebound).  Despite these studies, no formal standard has been adopted to classify a patient as having DSVL, SHVL, HVLS, SLVL, or rebounding viral load (RVL) patterns.  

Attempts to classify patient viral load states are further complicated by the challenges of analyzing viral load data.  Real-world viral load measurements are taken intermittently over time and may be absent due to a variety of factors (e.g. travel, social circumstances, non-adherence), leading to incomplete and irregularly spaced data points.  Differences in the sensitivity of the multiple assays used to measure viral loads result in different measures of ``undetectable" viral loads, further complicating analyses. Thus, there is a need for analytic techniques that can adjust for these details and classify viral load states. 

Machine learning methods can provide objective, unsupervised classification of patient clinical status\cite{KONONENKO200189}.  These methods begin by assigning a set of features to a patient (e.g. demographics, laboratory measurements, therapies) and then performing computational clustering to identify similar classes of patients.  Several studies have applied machine learning methods in HIV research\cite{dubey2016applications} to predict HIV viral load responses \cite{rosa2014insights} or CD4 T cell counts\cite{rodriguez2013predictions}, to distinguish between suppressed and viremic patients\cite{ramirez2016immunologic}, and to select therapeutic regimens\cite{parbhoo2017combining}. None, however, have used machine learning to create a standard classification for viral load status. 

To address these issues, we propose a set of unambiguous features, which combined (i.e. a feature vector), captures distinct dynamic patterns present in viral load measurements over time.   In addition, we have developed a novel centroid algorithm to cluster HIV positive subjects based on these patterns. Here we present the derivation of our machine learning method, and demonstrate its application to cluster 1,576 HIV patients with repeated VL measurements over a 5 year period.  We found that patient viral load measurements can be classified into five time-varying patterns, corresponding to clinically relevant states.  We note that the method and resulting categories can be used to standardize definitions of VL patterns across research studies.

\section*{Materials and Methods}
\subsection*{Human Subjects Protection}
This proposal was reviewed and approved by the University of Rochester Human Subjects Review Board (protocol number RSRB00068884). The analysis in this paper is presented in compliance with Center for Medicare Services (CMS) current cell size suppression policy\cite{cmsCell}.  Data were coded such that patients could not be identified directly in compliance with the Department of Health and Human Services Regulations for the Protection of Human Subjects (45 CFR 46.101(b)(4)). 

\subsection*{Study Data}
We obtained medical encounter data from all patients with an HIV diagnosis in the University of Rochester Medical Center's electronic medical record system (EMR) between 2011 and 2016. For each patient we have information on their age, gender, race, ethnicity, zip code, and coded procedures with associated ICD9 and ICD10 diagnosis codes. All patient viral load and CD4 measurements during the study period were also obtained.   There were a total of 1,892 patients with at least one viral load measured, with 1,576 of these patients having at least three viral load measurements. Viral load measurements $\leq$48 copies/mL, present as categorical values ``NEG", ``POS $<$ 20", or ``POS $<$ 48" were transformed into numerical values of 0, 20, and 48 respectively.

\subsection*{Data Availability}
The data is provided in 

\subsection*{Hardware and Software Specifications}
Analyses were performed on a Windows 8 server with Intel(R) Xeon(R) CPUs E5-2620 v2 @ 2.10GHz and 256GB of RAM. Python 2.7 was used for most data mining and machine learning under Spyder v.3 installed from Aanaconda2 (64-bit). The default packages available in Anaconda were used for analysis, including, but not limited to: NumPy, scikit-learn, SciPy, datetime, csv, math, Matplotlib, pip, operator, copy, random, and time.  Using pip we installed the webcolors and pydotplus packages for rendering a decision tree. Microsoft Excel was used to open, edit, and view flat files in .csv and .xlsx format. SQLite was used to store, query, and clean \textasciitilde 1.5 GB of data. Analytic code is available for download at: https://github.com/SamirRCHI/Viral\_Load\_Data\_Categorization.

\subsection*{Viral Load Analysis Methods}

Mathematical notations for this work are described in Table \ref{tab:notation}.

\begin{table}[!h]
\centering
\caption{
{\bf Notations.}}
\begin{tabular}{l|l}
Symbol          & Description                                                                                                                                                         \\ \hline
$N$             & The number of usable patients in the data. In our case 1576 patients.*                                                                                              \\
$p$             & Refers to a single patient.
\\
$VLM_p$         & The total number of viral load measurements patient $p$ has taken. \\
$\vv{VL}_p$      & All viral load counts of patient $p$ in order of time.
\\
$\vv{VL}_{p,i}$          & Refers to the $i^{th}$ viral load count of patient $p$ in $\vv{VL}_p$ , where $1 \leq i \leq VLM_p$.     \\
$\vv{t}_p$    & All temporal instances corresponding to $\vv{VL}_p$.
\\

$\vv{t}_{p,i}$           & Temporal instance of viral load $\vv{VL}_{p,i}$, where $1 \leq i \leq VLM_p$.                                                                                                                       \\
$max_{VL}$      & The maximum viral load for \textit{all} patients, ($10^7$).**                                                                                  \\
$min_{VL}$      & The minimum viral load for \textit{all} patients, ($0$).**   \\
$\circ$ & Hadamard Product - elemental-wise multiplication of arrays.\\
\hline 
\multicolumn{2}{l}{*This is after selecting for patients with $\geq 3$ measurements.}  \\
\multicolumn{2}{l}{**This value changes after transformation of the data.}  \\
\end{tabular}
\label{tab:notation}
\end{table}

\begin{figure}[!h] 
	\centering
    \includegraphics[width=0.9\linewidth]{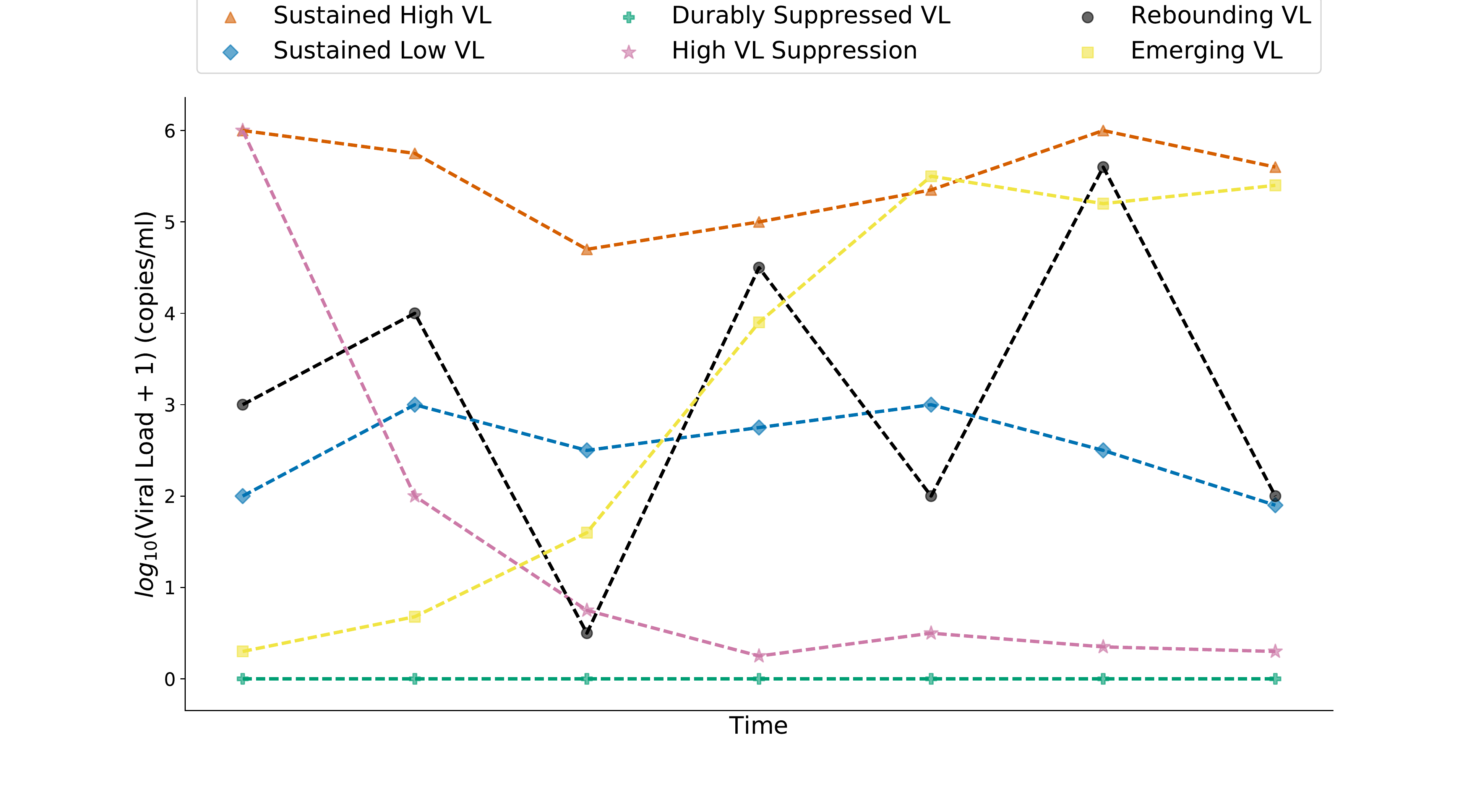}
    \caption{\textbf{Possible HIV Viral Load Patterns.} We provide a mock example of each type of viral load pattern that exist in the data.  Note that actual viral load patterns are noisier and may often be more difficult to distinguish.}
    \label{fig:general_cases}
\end{figure}

Based on temporal patterns of viral load described in the literature, viral load pair distribution of our data (\nameref{sfig:VLD}), and a further extensive investigation into the data, we hypothesized six potential temporal viral load patterns (Fig \ref{fig:general_cases}).

\begin{enumerate}
\item \textit{Durably Suppressed Viral Load} (DSVL): Having consistently suppressed their viral loads at or near the undetectable range.\\
\item \textit{Sustained Low Viral Load} (SLVL): Viral load counts which are constantly slightly higher than the undetectable range.\\
\item \textit{Sustained High Viral Load} (SHVL): Viral load counts which are constantly in a range considered high risk for HIV complications (e.g. opportunistic infections, malignancy).\\
\item \textit{High Viral Load Suppression} (HVLS): A viral load pattern in which the terminal portion of the curve has a negative slope and the terminal data point is in the low or suppressed range. This could have a few different speeds or styles of suppression - rapid, gradual, or slow.
\item \textit{Rebounding Viral Load} (RVL): A viral load pattern in which viral loads are unstable, with the measurement at one time step seemingly being independent of the next measurement.
\item \textit{Emerging Viral Load} (EVL): Having a steady, or rapid, emergence of high viral load while the first few measurements of viral load were suppressed. While we have found no mention of this type of pattern in the literature, and found that this pattern did not occur in our data set, VL data sets could contain this pattern.
\end{enumerate}

It is important to note that our definitions are pattern based, and do not explicitly select absolute viral load cutoff levels, nor selects a specific window, as previous reports have done \cite{greub2002intermittent,main,rose2015comparison,phillips2001hiv}. This has the advantage of allowing the absolute viral load levels, and critical time windows, to emerge from the analysis, and does not preclude incorporation of absolute levels (e.g. VLM$>$400) at a later stage into pattern specification.

\subsubsection*{Feature Vector Definition}

We next designed a feature vector to capture characteristics that would allow us to distinguish between viral load patterns.  Since viral load data is asynchronous and noisy, with variable numbers of data points for each subject, we argue that one or two viral load measurements are too few to accurately judge viral load behavior. Hence we restrict ourselves to using patients with $\geq$3 viral load measurements. In addition (\nameref{sfig:VLD}), viral load values at the lower limit of detection are a function of the specific assay used, and appear in our data set as 0, 20, and 48 copies/mL. Thus, plots of the $\log_{10}$ transformed data have discretely spaced values at the lower level of detection, although these values all capture the undetectable range of viral load. Additionally, we adjusted the data by $\log_{10}(VL+10)$ to avoid $\log_{10}(0)$. The addition of 10 to VL (instead of 1) is used to minimize the distance between the undetectable values: 0, 20, and 48 (copies/mL). Thus, in our notation, all the values related to viral load are assumed to have been adjusted to this measure. For example, $min_{VL} = \log_{10}(0+10) = 1$ and $max_{VL} = \log_{10}(10^7 + 10) \approx 7$.

Using the transformed VL data, we extract several relevant features of the VL measurements over time.  These features are used for machine learning classification of individual patient VL time series, and designed to distinguish patterns in VL change while minimizing the effects of noise, limited by the number of measurements. We do not limit feature extraction based on the total elapsed time of viral load measurements because the optimal time-point for determining viral load class is not well established. The attributes for feature extraction are: relative area of viral exposure, weighted recency reliability (wRR), adjusted maximal difference, and interquartile range (IQR).

\begin{enumerate}
\item \textit{Relative Area of Viral Exposure} (Area) - is defined as the area under the viral load curve relative to the total viral load area possible. Therefore the value of this feature will naturally be between 0 and 1. We choose a normalized, relative score, as the total time span between the first and last viral load measurement differs between patients. This feature is similar to finding the mean and median, except it is sensitive to the dimension of time, hence yielding more information. The feature is calculated by summing the area of each trapezoid created by each pair of viral load values, followed by dividing by the total possible area (Eq \ref{eq:relative_area}).

\begin{equation}
\dot{A}_p = \frac{\sum_{i=2}^{VLM_p} \frac{\vv{VL}_{p,i}  + \vv{VL}_{p,i-1}}{2}(\vv{t}_{p,i}  - \vv{t}_{p,i-1})}{(max_{VL} - min_{VL})(\vv{t}_{p,VLM_p} - \vv{t}_{p,1})}
\label{eq:relative_area}
\end{equation}

\item \textit{Weighted Recency Reliability} (wRR) - Due to viral load noise, the last viral load measurement is not necessarily an accurate reflection of the patient's overall, or most recent, viral load trend. For example, a patient may have a VL whose average slope is negative, indicating high viral load suppression over time (HVLS).  However, the last measurement may be slightly higher than the trend.  Thus heavily weighting the last measurement could lead to mis-classification as rebounding viral load (RVL).  To account for this, we calculate a weighted mean where the weight of the VL measurement increases with time. More specifically, the weight function follows an inverse square root function ($f(x) = \frac{1}{\sqrt{x}}$) rather than an inverse function ($g(x) = \frac{1}{x}$).  This has the advantage of avoiding rapid convergence of $g(x)$ to zero when time is measured in units of days (Eq \ref{eq:weight}). Weighted recency is then calculated as the dot product of the viral loads and the weights divided by the sum of the weights (Eq \ref{eq:wR}).

\begin{equation}
\vv{weight}_{p,i} = \frac{1}{\sqrt{\vv{t}_{p,VLM_p}-\vv{t}_{p,i}}+1}
\label{eq:weight}
\end{equation}

\begin{equation}
wR_p = \frac{\vv{weight}_p \bullet \vv{VL}_{p}}{\sum_{i=1}^{VLM_p} \vv{weight}_{p,i}}
\label{eq:wR}
\end{equation}

We were also interested in how reliable $wR$ is as a representation of the patient's viral load trend. To this end, we calculated the absolute deviations from the viral load measurements to $wR$ (Eq \ref{eq:dev}). Rather than averaging the deviations, we take the median to reduce the effects of outliers and call this our weighted recency reliability measure (Eq \ref{eq:wRR}). We take the inverse to force the range of the result to be between [0,1]; a property made to use in our next proposed feature, adjusted maximal difference.

\begin{equation}
\vv{dev}_{p,i} = |wR_p - \vv{VL}_{p,i}|
\label{eq:dev}
\end{equation}

\begin{equation}
wRR_p = \frac{1}{\textrm{median}(\vv{dev}_p)+1}
\label{eq:wRR}
\end{equation}

\item \textit{Adjusted Maximal Difference} (Adj MD) - We used a time-independent method to measure the final change in viral load by determining maximal VL difference defined as the difference between the ``peak'' and last VL measurements. To distinguish between viral load suppression or emergence, we calculate the ``peak'' as the maximum of the absolute deviations (Eq \ref{eq:dev}) and retain the sign of the result. We expected the positive scores to effectively isolate the EVL group, however, we instead found that retaining the positive (emergent) scores lead to mis-categorization of SHVL and RVL groups without clearly identifying EVL patterns. This, along with other investigation into the data, led us to conclude that the EVL pattern does not exist in our data, but we refrain to make generalizations to all healthcare facilities. Be that as it may, we force (ground) the positive scores down to zero for proper labeling of SHVL and RVL (Eq \ref{eq:grnd}).

Due to the varying nature in viral load measurements, we are hesitant to use the final viral load measurement as a means of judging suppression. Thus we propose to use $wR$ instead. To reduce the effects of rebounding patients being falsely labeled as suppressed patients, we multiply our result by $wRR$ - as rebounding patients are expected to have a low score in the range [0,1]. The maximal difference is necessary in order to ensure that the suppression type of viral load patterns are classified appropriately (Eq \ref{eq:rel_max_diff}).

\begin{equation}
\textrm{grnd}(x) = 
\begin{cases}
	-1 & x < 0 \\
    0 & x \geq 0
\end{cases}
\label{eq:grnd}
\end{equation}

\begin{equation}
\check{D}_p = \textrm{grnd}(wR_p - \vv{VL}_{p,\textrm{argmax}(\vv{dev}_p)})*\textrm{max}(\vv{dev}_p) * wRR_p
\label{eq:rel_max_diff}
\end{equation}

\item \textit{Interquartile Range} (IQR) - This feature is added to further segregate the rebounding patients and follows the standard interquartile range calculation (Eq \ref{eq:IQR}).

\begin{equation}
IQR_p = \textrm{Q}_3(\vv{VL}_p) - \textrm{Q}_1(\vv{VL}_p)
\label{eq:IQR}
\end{equation}

\end{enumerate}

\subsection*{Analytic Terminology}
Here we formally define keywords which will appear in the analysis: Let \textit{Feature extraction} be the process of determining the values $\dot{A}$, $wRR$, $\check{D}$, and $IQR$ from a set of patients (using their viral load patterns) with the formulations given above. Then a \textit{feature vector} ($\vec{F}_p$) contains the values $\dot{A}_p$, $wRR_p$, $\check{D}_p$, and $IQR_p$ extracted from patient $p$'s viral load pattern. The words \textit{sample} or \textit{point} are also used here interchangeably. The term \textit{feature} ($F$) can be thought of as a column vector for all patients in the dataset consisting of the four attributes: $F_{\dot{A}}$, $F_{wRR}$, $F_{\check{D}}$, and $F_{IQR}$ (\nameref{sfig:outline}A). Finally, the terms \textit{label assignment}, \textit{VL pattern membership assignment}, \textit{patient categorization}, and \textit{prediction}, all refer to the same principle: To assign the most appropriate label which characterizes the viral load pattern of a patient. However, while the principle is the same, the method of assigning such an appropriate label differs depending on the categorization method, or the learning method, being used.

\section*{Results}

\subsection*{Feature Extraction and Normalization}

We began by transforming viral load data by min-max normalization \cite{han2011data} to equally weight the temporal features of the VL series (Eq \ref{eq:normalize}). That is, we normalize the features, $F$, to a range between [0, 1] using equation \ref{eq:normalize} where $F^* = f(F)$ (\nameref{sfig:outline}B.1).  
\begin{equation}
F^* = f(F) = \frac{F - \min F}{\max F - \min F}
\label{eq:normalize}
\end{equation} 
Next, we examined each of the 4 features for all patients with $\geq 3$ viral load measurements ($N = 1576$ patients) (\nameref{sfig:outline}A), which did not exhibit distinct bi-variate clustering (\nameref{sfig:bivariate}). A correlation coefficient analysis between the features (Table \ref{stab:correlationcoefficient}) reveal that the Adj MD feature appears to be linearly independent to Area and wRR. There does seem to be some linear dependence between IQR and Adj MD, and Area to wRR and IQR. As expected, the largest linear dependency appears to be between wRR and IQR. These results suggest the separation between viral load patterns will be most noticeable between the Area and the Adj MD features - as we designed them to be. Also, although Adj MD is dependent upon wRR, we find that their correlation coefficient is very low (0.033) as they try to capture two very different phenomenon.

\subsection*{Hierarchical Clustering}

We then performed hierarchical clustering of the individual subjects using a Euclidean distance metric and the Ward's criterion\cite{punj1983cluster} to minimize the total within-cluster variance, revealing a clear separation into 5 distinct patient groups (\nameref{sfig:outline}C).

\begin{figure}[!h] 
	\centering
    \includegraphics[width=1\linewidth]{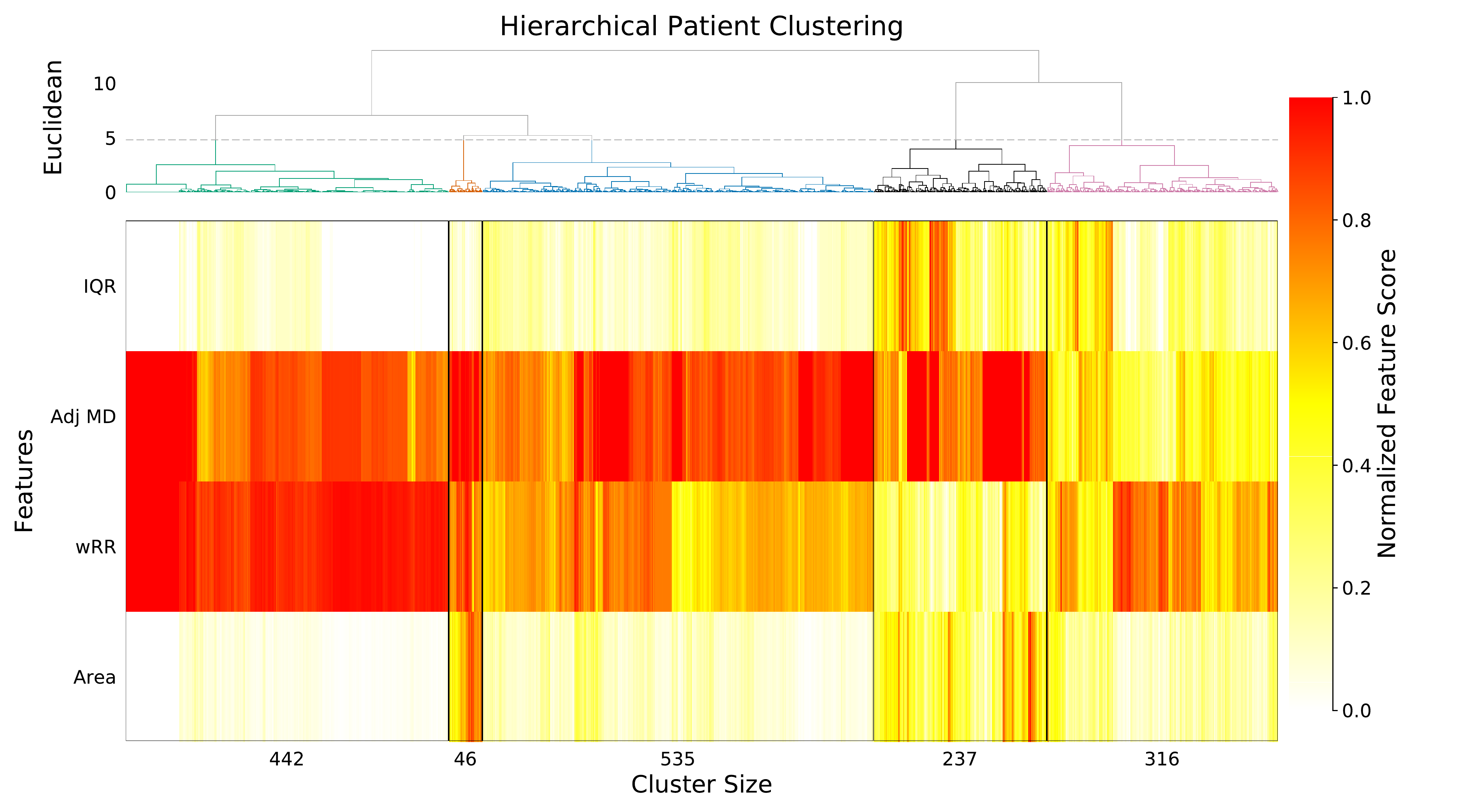}
    \caption{\textbf{Dendrogram of hierarchically clustered patients}. Clustered using the euclidean distance along with the ward method. Numbers on the bottom axis show number of patients in each cluster.  The corresponding viral load pattern plots can be found in Fig~\ref{fig:VL_patterns}. Colors are derived from a 7-color palette optimally designed for color-blind individuals \cite{wong2011points}.}
\label{fig:HPC}
\end{figure}

\begin{figure}[!h] 
	\centering
    \includegraphics[width=1\linewidth]{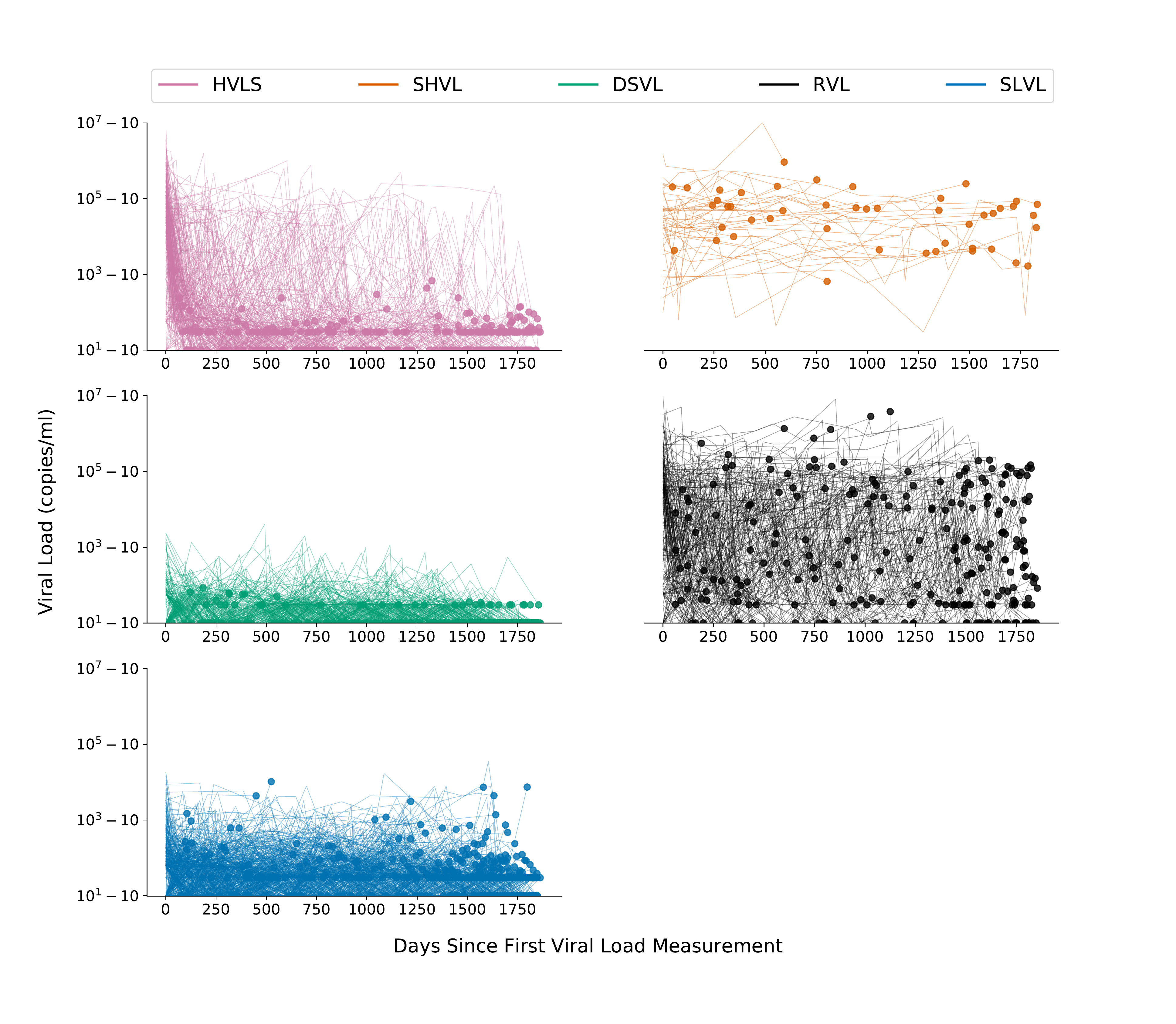}
    \caption{\textbf{Extracted patient viral load patterns.} For each cluster categorization of the patient from Fig \ref{fig:HPC}, the days since first viral load measurement are plotted against the viral load counts. The points on the plots indicate the last viral load measurement.}
    \label{fig:VL_patterns}
\end{figure}

\begin{figure}[!h] 
	\centering
    \includegraphics[width=1\linewidth]{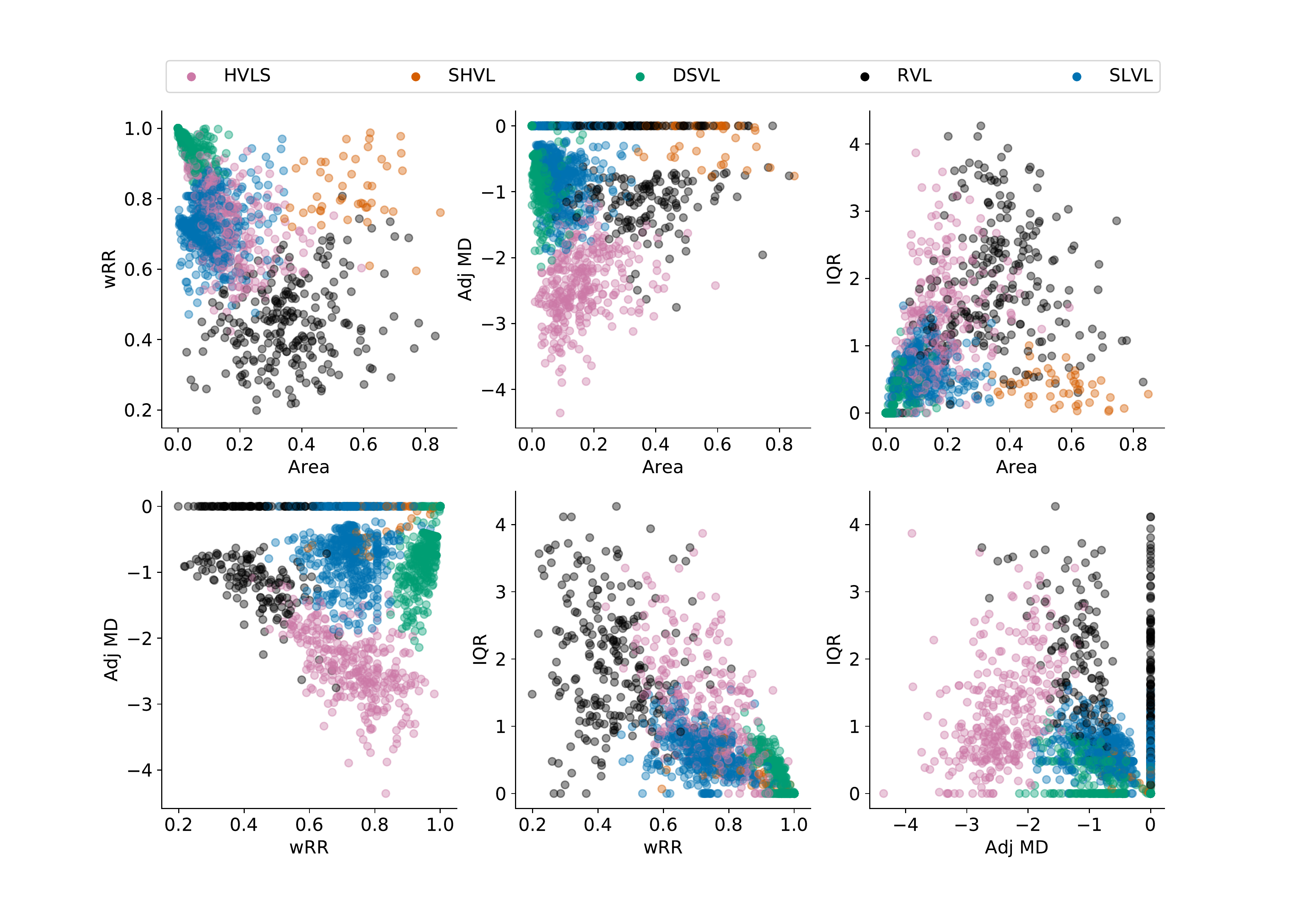}
    \caption{{\bf Feature segregation from hierarchical clustering.} Shows the same scatter plot from Fig S2 but with each patient colored corresponding to the results from the hierarchical clustering in Fig \ref{fig:HPC}. The artificial line of points is a result of the grounding function used in Adj MD.}
    \label{fig:bivariateclustering}
\end{figure}

From Fig \ref{fig:HPC} and Fig \ref{fig:VL_patterns}, we find that the bluish green cluster (n=442) corresponds to the DSVL patient group, showing the lowest viral loads amongst the clusters identified and the highest weighted recency reliability. The patients in the vermillion-red cluster (n=46) correspond to the SHVL group, exhibiting the highest relative area and very low IQR. Compared to the DSVL cluster, the blue cluster (n=535) has slightly greater area and IQR with a significant difference in the weighted recency reliability. Using this information, along with the general patterns shown by Fig \ref{fig:VL_patterns}, we identify this as being the SLVL group. The algorithm also identifies the black (n=237) and redish purple (n=316) clusters, which appear to correspond to the RVL and HVLS classes respectively. The RVL cluster has a low weighted recency reliability and high IQR. In contrast, the HVLS cluster has a lower area, higher weighted recency reliability and most importantly very low adjusted maximal difference (Fig \ref{fig:bivariateclustering}).

Upon examining Fig \ref{fig:VL_patterns}, we also find that the viral load patterns seem to be similar within cluster, and dissimilar between clusters. Interestingly, in each extracted cluster, we find that there are a few patients whose last viral load measurement was taken around 1,827 days - which is equivalent to the full span of five years that our patient's viral loads have been monitored. This suggests that these clusters don't disappear after some elapsed time, but rather each type of pattern can be found at virtually any time point. 

In regards to the HVLS group, we find large spikes in the middle of their time-series. We hypothesize that this may be due to the asynchronous timing of measurements between subjects, the natural variation in biological responses between the groups, or patient variability in adherence to therapy.  This observation also reflects one limitation of asynchronous outcomes data sampling without a ``completion" endpoint, a characteristic of most prospective, randomized clinical trials.  If measurements were stopped at one of the spikes, the adjusted maximal difference feature may be weighted in the favor of being classified as a patient whose viral load is rebounding. This phenomenon could be viewed with two different perspectives: First, this may indicate that the some patients labeled as having suppressed their viral loads should have been labeled as having rebounding viral loads. Or second, this changing in class labels show that these features do not restrict a patient to forever be restricted to one category. Rather - depending on biological or therapeutic responses, the patient may now belong to a different category all-together.

\subsection*{Comparison with Existing Categorization Methods}

\begin{figure}[!h] 
	\centering
    \includegraphics[width=1\linewidth]{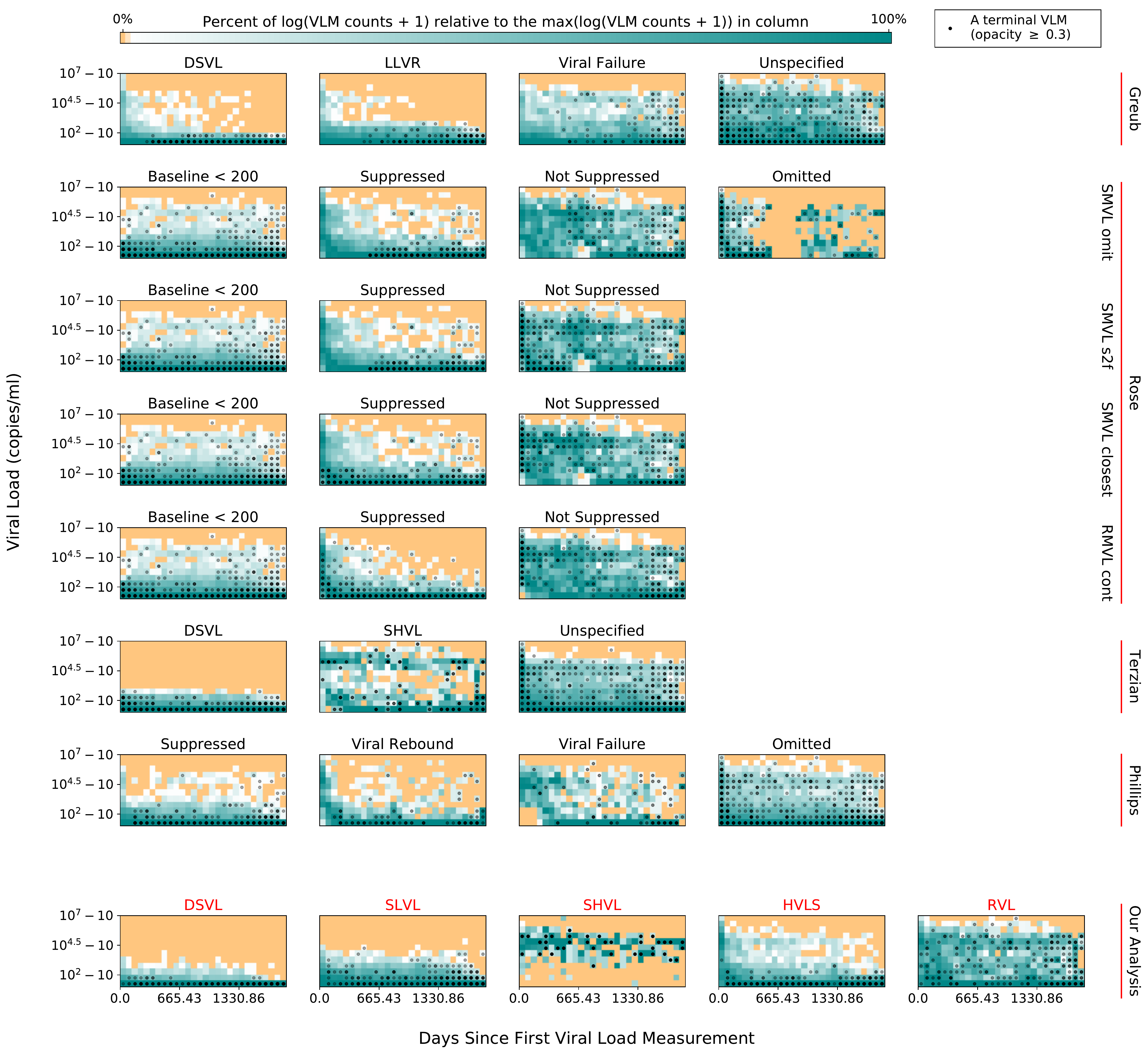}
    \caption{\textbf{Comparison of Patient Categories with Existing Methods.} Shows a 2D binning of VLM counts for every patient category. Each row uses a different categorization method in labeling the patients in our dataset; the method name is located to the right of the row. The title of each subplot is the label assigned by the indicated method. The columns of each 2D bin are normalized based on the maximum number of logged VLM counts in the column (add by 1 to avoid $\log(0)$). Bin color for a count of 0 is copper, and the rest of the bin colors ranges from white to teal (the maximum of the $\log(VLM\ counts)$ in the column of the bin). The black dots represent the last viral load measurement taken by the patient (opacity $\geq$ 0.3; 2D bins have variable opacity for the dots). The bottom row is our analysis, which is the same as Fig \ref{fig:VL_patterns}, but represented as a 2D bin.}
    \label{fig:comparison}
\end{figure}

Visually, we find that the SLVL group detected by our method is very similar to the LLVR group defined by Greub et. al. (Fig \ref{fig:comparison}). Furthermore, it appears that the methods trying to capture SHVL, viral rebound, and viral failure patients did not succeed as well as the identification of SHVL and RVL patients in our method. RMVL repeat continuous visually appears to have performed very well in identifying patients whom have suppressed their viral loads. However, we argue that our analysis performs perhaps slightly better in identifying the suppression group (HVLS), as we find that the last viral load measurement for each patient (black dots in Fig \ref{fig:comparison}) are consistently low using our method. The existing methods may not have performed as well because they rely on a window or a consecutive pair measure which may be too subjective for assigning VL pattern membership. Furthermore, based on Fig \ref{fig:comparison}, Rose et. al.'s assumption, that patients with a baseline viral load less than 200 have consistently low viral variation, may be subject to scrutiny. Lastly we wish to emphasize that while some of these existing categorization methods are successful in identifying a specific group of patients, our method is unique as it attempts to give definitive group labels to each viral load pattern.

\subsection*{Classification Stability}

To address the potential issue of misclassification, and to assess the changing nature of viral load pattern membership assignment, we performed a time-varying classification sensitivity analysis. First, we train a supervised machine learning method on the results from our original classification. Then, to determine the classification stability of our results, we iteratively determine membership assignment given only partial viral load data for every patient. Our objective is to pair the feature vector with a supervised learning method that has high classification stability in the setting of incomplete, time varying, VL data.  Additionally, the method should be easy to apply by clinical researchers to assign viral load pattern membership to their patients for cohort studies. Hence, easy interpretation and reconstruction of the model is ideal.

\subsubsection*{Centroid Summarization}

Prior to performing partial viral load membership assignment, we initially consider several common supervised learning methods for learning on our initial classification results: k-nearest neighbors (k=5,7,9) \cite{peterson2009k}, support vector machine (SVM), decision tree, AdaBoost, and random forests. We tested the performance of these methods with leave-one-out cross-validation (LOOCV) on our initial classification results, and by calculating the $F_1$ scores of the prediction of each category.  The $F_1$ score is the harmonic mean of precision and recall \cite{han2011data} defined as equation \ref{eq:F}. The results of each are superlative with the strange exception of AdaBoost (Table \ref{tab:2}).

\[
precision = \frac{True\ Positive}{True\ Positive + False\ Positive}
\]
\[
recall = \frac{True\ Positive}{Positive} 
\]
\begin{equation}
F_1 = 2 * \frac{precision*recall}{precision + recall}
\label{eq:F}
\end{equation}
\begin{table}[htbp]
  \caption{\textbf{$\boldsymbol{F_1}$ prediction scores using LOOCV.}}
    \begin{tabular}{lrrrrr|r}
    \toprule
    Group: & \multicolumn{1}{l}{\textcolor{DSVL}{DSVL}} & \multicolumn{1}{r}{\textcolor{SLVL}{SLVL}} & \multicolumn{1}{r}{\textcolor{SHVL}{SHVL}} & \multicolumn{1}{r}{\textcolor{HVLS}{HVLS}} & \multicolumn{1}{r|}{\textcolor{RVL}{RVL}} & \multicolumn{1}{r}{Average} \\
    Patients: & 442 & 535 & 46 &316 & 237 & $F_1$ Score \\
    \midrule
    kNN,k=5 & 0.9966 & 0.9925 & 0.9677 & 0.9889 & 0.9810 & 0.9853 \\
    kNN,k=9 & 0.9943 & 0.9907 & 0.9583 & 0.9841 & 0.9725 & 0.9800 \\
    kNN,k=7 & 0.9943 & 0.9897 & 0.9362 & 0.9873 & 0.9746 & 0.9764 \\
    Random Forest & 0.9909 & 0.9841 & 0.9556 & 0.9685 & 0.9645 & 0.9727 \\
    Decision Tree & 0.9898 & 0.9795 & 0.9670 & 0.9512 & 0.9432 & 0.9661 \\
    SVM   & 0.9955 & 0.9833 & 0.9111 & 0.9666 & 0.9387 & 0.9590 \\
    \textit{Polyhedron} & 0.9727 & 0.9474 & 0.9011 & 0.8985 & 0.9109 & 0.9261 \\
    \textit{Bounding Box} & 0.9865 & 0.9630 & 0.8764 & 0.9038 & 0.8614 & 0.9182 \\
    \textit{Push and Pull} & 0.9737 & 0.9347 & 0.8842 & 0.9027 & 0.8767 & 0.9144 \\
    \textit{Best Rep.} & 0.9589 & 0.9280 & 0.9011 & 0.8598 & 0.8923 & 0.9080 \\
    \textit{Mean} & 0.9401 & 0.9004 & 0.9072 & 0.8212 & 0.8968 & 0.8931 \\
    \textit{Smallest Disk} & 0.9627 & 0.9017 & 0.8889 & 0.8246 & 0.8717 & 0.8899 \\
    \textit{Median} & 0.9271 & 0.8882 & 0.9167 & 0.7967 & 0.8953 & 0.8848 \\
    AdaBoost & 0.9227 & 0.8248 & 0.5797 & 0.5033 & 0.6475 & 0.6956 \\
    \bottomrule
    \end{tabular}%
  \label{tab:2}%
\end{table}%

However, these supervised learning techniques defeat our objective of offering a simple model which clinicians and researchers may effortlessly reconstruct in the programming language of their choice. More specifically, k-Nearest Neighbors (kNN) is inherently ``fuzzy" \cite{keller1985fuzzy}, meaning that prediction on new samples is dependent upon the entire training set, and so the learned patterns are ambiguous. While SVM offers a simpler model, the results can be non-intuitive for clinicians. Finally, although Decision Trees offer the best interpretability, overly complex trees may be generated, as it did in our case (\nameref{sfig:DecisionTree}). Thus, we next set out to develop a more robust unsupervised method with improved interpretability.

We noted that, similar to k-Means, the hierarchical clusters naturally form non-intersecting clouds in $n$ dimensions \cite{rohlf1970adaptive} (i.e. the non-convex hulls tend not to intersect). The cloud centroids can be algorithmically identified to yield a summarization of the clusters, drawing similarities with the cluster summarization involved in the BIRCH algorithm \cite{han2011data}. Using this property, we can calculate the radius of the cluster as the farthest intra-cluster sample from the cluster centroid to reduce misrepresentation of distant member objects - rather than calculating the radius as the average distance between the centroid to the object members (as the BIRCH algorithm does). We propose a radius-based prediction that operates as follows: Let $c_i$ be the $i$th cluster center with corresponding radius $r_i$, then for a new sample $s$ choose its predicted cluster membership $j$ such that $\frac{||s-c_j||_2}{r_j}$ is a minimum. We call this \textit{radial normalization} (RN) classification.

The definition of a center of points varies based on the shape of the data (e.g. Gaussian n-dimensional spheres, irregular density distributions) and on the algorithm used for identification. We therefore compared seven different methods to identify cluster ``centers" to assess classification stability: multidimensional mean, multidimensional median, best representative center, bounding box method, smallest disk method, polyhedral center, and a novel force directed ``push and pull" method inspired by force-directed graph drawing using Fruchterman-Reingold's algorithm \cite{kobourov2012spring, frurhineGraph} (see Supplementary Material for a detailed explanation of each method).  Force directed clustering methods maximize inter-cluster center distances while minimizing intra-cluster distance, and are the basis for modularity clustering in graph theory\cite{PhysRevE.79.026102}.   

The results for our evaluation of the centroid methods (CM) using RN (CMuRN) are italicized in Table~\ref{tab:2}. BIRCH defines its centroid as the average of all the points (multidimensional mean) \cite{han2011data} which scored lower than four other centroid methods. From amongst these CM, the polyhedral CMuRN scored the highest in terms of average $F_1$, and although this method did not outperform the more intricate supervised machine learning methods, the performance is nevertheless exceptional for its interpretability. We argue that any CMuRN is highly interpretable because the entire model can be summarized concisely (Table \ref{tab:Centroidres}) and understood clearly. For example, we learn from Table \ref{tab:Centroidres} that a characteristic of DSVL patients is that their relative area of severity and IQR are very low (0.0901 and 0.0983 respectively in the normalized region) and level of their viral loads are very consistent, judged by its weighted recency reliability (0.905 in the normalized region), with small flexibility in these values - as indicated by the radius. For its performance and interpretability, we subsequently use the polyhedral CMuRN to measure classification stability in the next section.

\begin{table}[htbp]
  \caption{{\bf  Learned centroids and radii from polyhedral CMuRN}}
    \begin{tabular}{lllllll|l}
    \toprule
          &   & Area  & wRR   & Adj MD & IQR   &  & \textit{Radius} \\
\hline
\multicolumn{1}{c}{\multirow{2}[2]{*}{\begin{sideways}LT\end{sideways}}} & $m_F$     & \multicolumn{1}{r}{1.179} & \multicolumn{1}{r}{1.2476} & \multicolumn{1}{r}{0.2295} & \multicolumn{1}{r}{0.2342} &       &  \\
& $b_F$     & \multicolumn{1}{r}{0} & \multicolumn{1}{r}{-0.2476} & \multicolumn{1}{r}{1} & \multicolumn{1}{r}{0} &       &  \\
\midrule
\multicolumn{1}{c}{\multirow{5}[2]{*}{\begin{sideways}LT(Centers)\end{sideways}}} & \textcolor{DSVL}{DSVL} & \multicolumn{1}{r}{0.0901} & \multicolumn{1}{r}{0.905} & \multicolumn{1}{r}{0.7803} & \multicolumn{1}{r}{0.0983} &       & \multicolumn{1}{r}{0.2946} \\
& \textcolor{SLVL}{SLVL} & \multicolumn{1}{r}{0.1848} & \multicolumn{1}{r}{0.638} & \multicolumn{1}{r}{0.848} & \multicolumn{1}{r}{0.1573} &       & \multicolumn{1}{r}{0.4124} \\
& \textcolor{SHVL}{SHVL} & \multicolumn{1}{r}{0.6696} & \multicolumn{1}{r}{0.7293} & \multicolumn{1}{r}{0.9302} & \multicolumn{1}{r}{0.0945} &       & \multicolumn{1}{r}{0.3518} \\
& \textcolor{HVLS}{HVLS} & \multicolumn{1}{r}{0.248} & \multicolumn{1}{r}{0.6498} & \multicolumn{1}{r}{0.4293} & \multicolumn{1}{r}{0.3758} &       & \multicolumn{1}{r}{0.6354} \\
& \textcolor{RVL}{RVL} & \multicolumn{1}{r}{0.4947} & \multicolumn{1}{r}{0.3513} & \multicolumn{1}{r}{0.8111} & \multicolumn{1}{r}{0.4914} &       & \multicolumn{1}{r}{0.7319} \\
\bottomrule
    \multicolumn{8}{l}{Linear Transformation (LT) determined by equation \ref{eq:LT}.} \\
    \end{tabular}%
  \label{tab:Centroidres}%
\end{table}%

\subsubsection*{Membership Assignment on Partially Retained VL Patterns}

To determine the classification stability, we iteratively assign VL pattern membership using only partial viral load data for every patient. First, we extract the feature vector $\vv{F}$ from the partially retained data (\nameref{sfig:outline}G.2).  We then transform $\vv{F}$ to the same normalized space the polyhedral CM was trained on. Notice the min-max normalization (Eq \ref{eq:normalize}) is a linear transformation (LT; Eq \ref{eq:LT}), hence the transformation is performed by applying $LT$ onto $\vv{F}$ (i.e. $LT(\vv{F})$; \nameref{sfig:outline}H.2), with the parameters for the LT given in Table \ref{tab:Centroidres}.

\[
m_F = \frac{1}{\max F - \min F} \implies \vv{m} = [m_{F_{\dot{A}}} \quad  m_{F_{wRR}} \quad m_{F_{\check{D}}} \quad m_{F_{IQR}}]
\]
\[
b_F = \frac{-\min F}{\max F - \min F} \implies \vv{b} = [b_{F_{\dot{A}}} \quad  b_{F_{wRR}} \quad b_{F_{\check{D}}} \quad b_{F_{IQR}}]
\]
\begin{equation}
LT(\vv{F}) = \vv{m} \circ \vv{F} + \vv{b}
\label{eq:LT}
\end{equation}

\begin{figure}[!h] 
	\centering
    \includegraphics[width=1\linewidth]{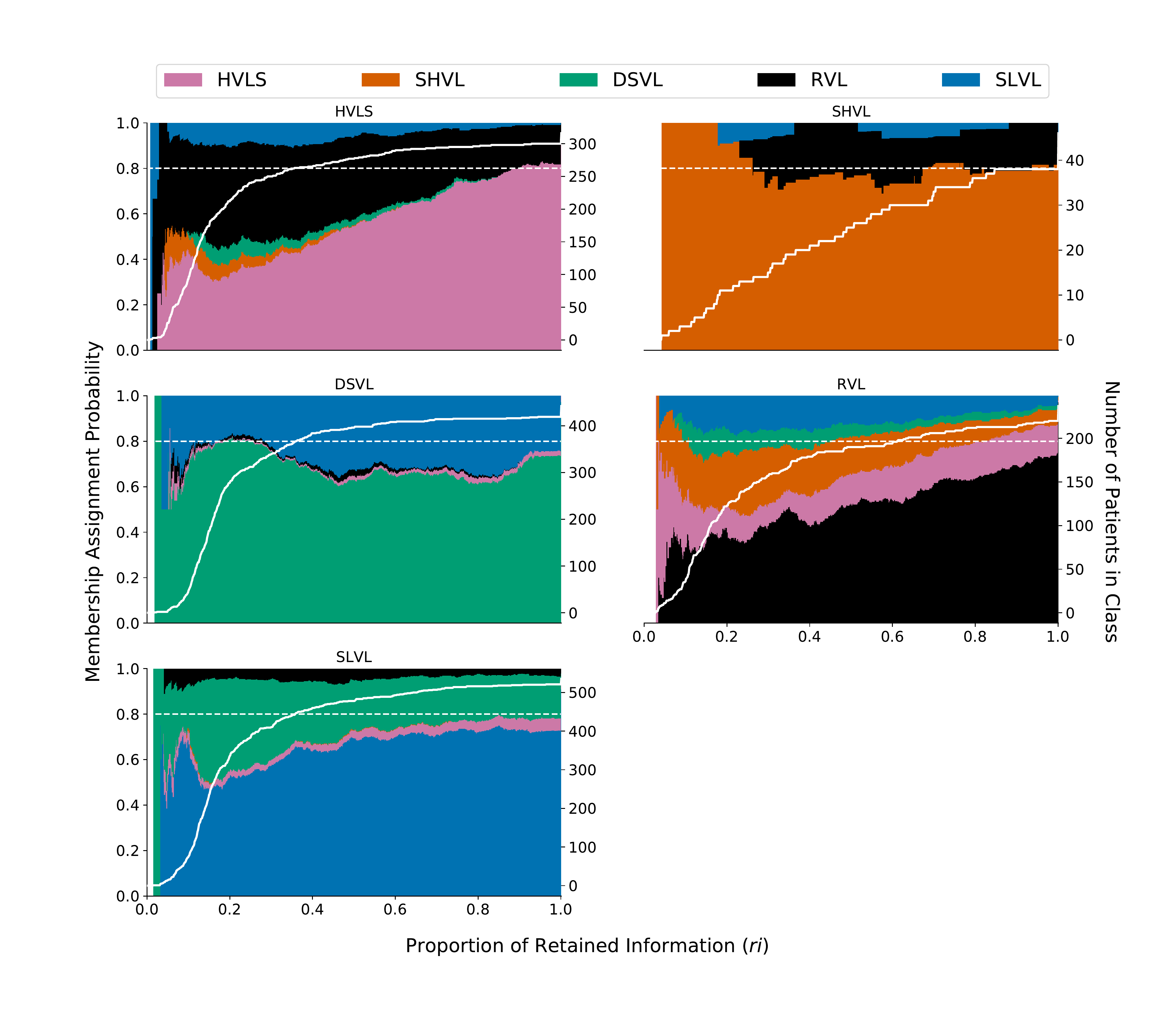}
    \caption{\textbf{Polyhedral CMuRN Classification Stability.} Classification stability results of polyhedral CMuRN learned on 100\% retained data. The title of each plot corresponds to which category of patients are being tested. The dashed white line is the 80\% membership assignment probability line, included for reference. The solid white line represents the number of patients included in the prediction at the point of retained viral load data (corresponding to the right axis) which changes as a result of the $r \geq 3$ restriction.}
    \label{fig:CCS}
\end{figure}

With the transformed feature vector, we assign viral load pattern membership using \textit{radial normalization} classification (\nameref{sfig:outline}I), where the centroid algorithm is trained on 100\% retained data (Table \ref{tab:Centroidres}). We repeat this procedure starting at 0\% retained information ($ri$) and progressively increasing to 100\% by a factor of 0.1\% for each patient (\nameref{sfig:outline}E.1-E.3). We record the change in patient viral load pattern membership upon each retained information instance, which we designate as \textit{membership assignment probability} (Fig \ref{fig:CCS}).

Because the patient will not have a measurement exactly at each instance of $ri$, we selected the nearest cutoff instance, $r$, such that $t_{p,r} \leq ri*\vv{t}_{p,VLM_p}$ (\nameref{sfig:outline}F). The feature vector was then extracted from $\vv{t}_{p,i}$ and $\vv{VL}_{p,i}$ where $i = 1, \ldots, r$. We also added the restriction $r >= 3$ to adhere to the minimum of three measurements requirement. If $r$ failed to meet this requirement, the patient was dropped from the centroid prediction instance $ri$ (\nameref{sfig:outline}G.1). This restriction explains why, if Fig \ref{fig:CCS} is examined closely, the classes have small gaps between the point where $ri = 0$ and the $ri$ value for which calculating membership assignments start.

The true positive probabilities associated with the DSVL, SLVL, and SHVL clusters are generally high (close to the range of 80\%) despite the amount of retained viral load data (Fig \ref{fig:CCS}). The true positive probability associated with the HVLS cluster is initially very low, but progressively reaches 80\% within 85\% of the retained viral load data. This type of progression is likely because the viral load suppression class is time dependent and thus VL time series taken early, or later, with respect to treatment initiation would be expected to differ based on the full treatment response. Similarly we find the RVL group's true viral load membership assignment increases with an increasing fraction retained information.

\begin{figure}[!h] 
	\centering
    \includegraphics[width=1\linewidth]{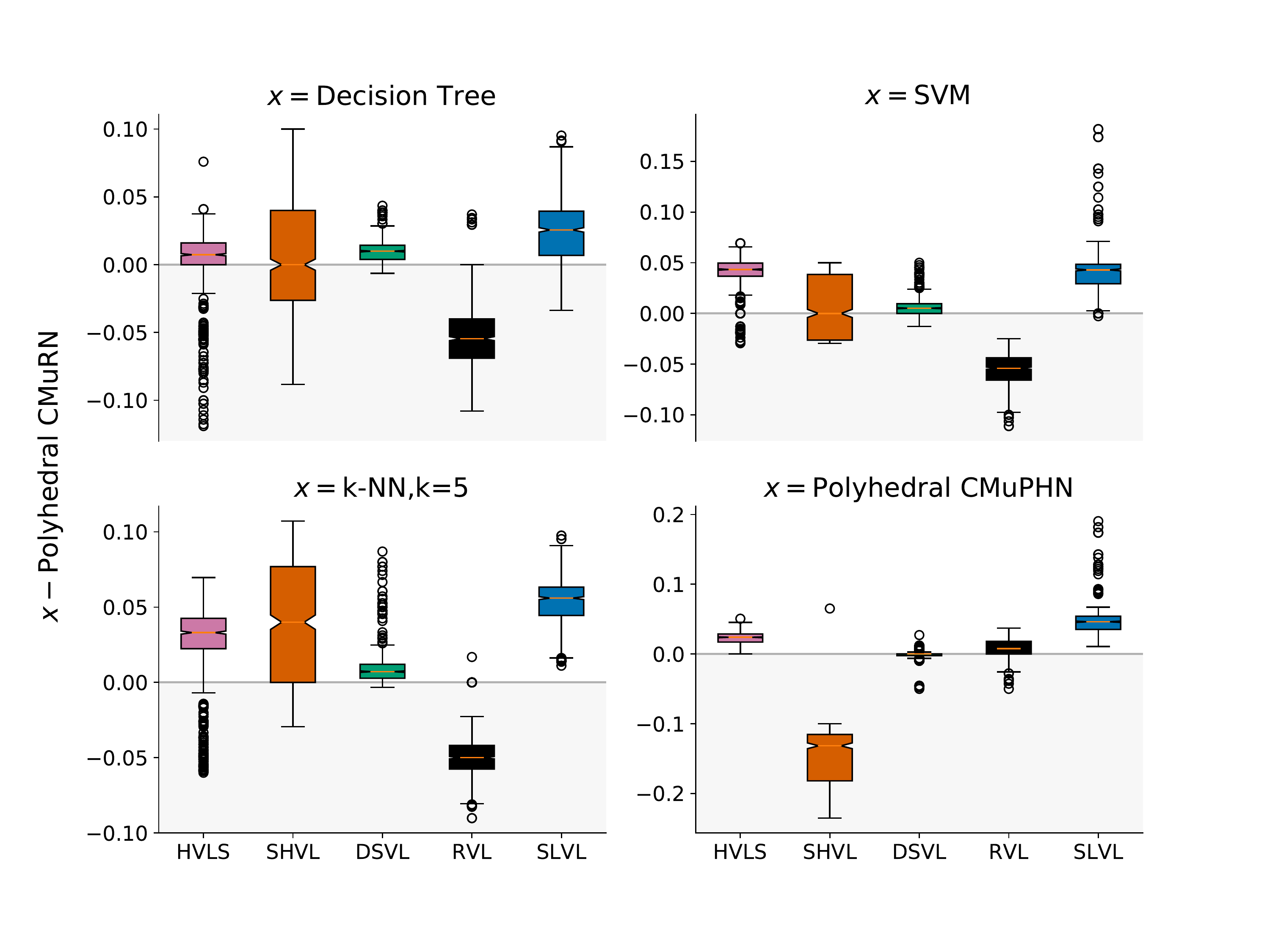}
    \caption{\textbf{Distributions of Classification Stability Performance Differences.} A comparison of the polyhedral CMuRN against the decision tree, SVM, kNN (k=5) and polyhedral CMuPHN algorithms in terms of true positive classification stability performance. Negative results (light gray region) imply that the Polyhedral CMuRN performed better than the comparison algorithm. A Wilcoxon signed-rank test yielded ($p < 0.001$) for each difference distribution.}
    \label{fig:Distributions}
\end{figure}

Next, we compared the classification stability results of the polyhedral CMuRN with the decision tree, SVM, and kNN algorithms. While differences in classification stability appear to be negligible to the naked eye (Fig S6-8), a Wilcoxon signed-rank test\cite{wilcoxon1945individual} between each true positive viral load pattern membership reveal that the difference in performance of each method is statistically significant for all viral load patterns ($p < 0.001$). Note that the test only includes the $ri$ instances where the number of patients are $\geq 20$. The distribution of these differences (Fig \ref{fig:Distributions}) reveal that the polyhedral CMuRN outperforms all other algorithms in terms of RVL assignment by $\approx$5\% (Table \ref{stab:quartiles}). Although the differences in performance may be statistically significant, the level of difference may not be clinically relevant.

\subsection*{Projected Hyperplane Normalization}

We were initially perplexed to find the smallest disk scoring low in Table \ref{tab:2} because this method attempts to find the center by minimizing the radius, hence one may speculate that it should score the highest from amongst the centroid methods using RN. However, the result led us to recognize that one potential drawback of RN is in assuming the cloud of points is spherical. This assumption can be relaxed by determining the set of hyperplanes which make up the convex hull of the cloud. In this way we may normalize $||s-c_j||_2$ by the distance to the nearest projected hyperplane in the convex hull. Such normalization improves the centroid results (Table \ref{tab:4}), but at the expense of model interpretability, which may not be desirable in a clinically applied method. Furthermore we find the classification stability results using projected hyperplane normalization (PHN) are slightly better than radial normalization (\nameref{sfig:PCShn}), with a seemingly negligible exception of DSVL and a major exception of SHVL performance dropping between 11 and 18 percent (Table \ref{stab:quartiles}), which may be a result of RVL and SHVL polyhedrons having a greater overlapping volume ratio than would their spheres. Thus this method also has its drawbacks. Performing the Wilcoxon signed-rank test between the polyhedral CMuPHN and polyhedral CMuRN also revealed statistical significance in the performance difference for each true positive viral load membership ($p < 0.001$) \cite{wilcoxon1945individual}.

We see two limitations to projected hyperplane normalization. First, the projection only works if the center is inside the convex hull. This leads to an issue with using the median and best representative methods because it is possible they may choose a center lying on the border of the convex hull. Even worse for this prediction method is using push and pull, which may detect a center outside the convex hull. Due to this, these three methods are not mentioned in the results of Table \ref{tab:4}. Second, finding the convex hull is computationally expensive and, depending on the algorithm, the accuracy of the convex hull becomes unwieldy with $>$15 dimensions \cite{avis1997good}.

\begin{table}[htbp]
  \centering
  \caption{\textbf{Projected Hyperplane Normalization $F_1$ scores using LOOCV }}
    \begin{tabular}{lrrrrr|r}
    \toprule
       Group:   & \multicolumn{1}{r}{\textcolor{DSVL}{DSVL}} & \multicolumn{1}{r}{\textcolor{SLVL}{SLVL}} & \multicolumn{1}{r}{\textcolor{SHVL}{SHVL}} & \multicolumn{1}{r}{\textcolor{HVLS}{HVLS}} & \multicolumn{1}{r|}{\textcolor{RVL}{RVL}} & Average\\
    Patients: & 442   & 535   & 46    & 316   & \multicolumn{1}{r|}{237} &  $F_1$ Score \\
    \midrule
    Polyhedron & 0.9851 & 0.9747 & 0.9048 & 0.9575 & 0.9495 & 0.95432 \\
    Smallest Disk & 0.9851 & 0.9776 & 0.9048 & 0.9542 & 0.9433 & 0.9530 \\
    Bounding Box & 0.9792 & 0.9718 & 0.9176 & 0.9428 & 0.9512 & 0.95252 \\
    Mean  & 0.9886 & 0.9766 & 0.878 & 0.9573 & 0.9476 & 0.94962 \\
    \bottomrule
    \end{tabular}%
  \label{tab:4}%
\end{table}%

\section*{Discussion}

Researchers have previously performed HIV population case studies using differing  schema to classify VL patterns \cite{main,yehia2012sustained,rose2015comparison,phillips2001hiv}. Our work is unique as it suggests a method for standardizing the classification of VL patterns using a set of optimally segregating features. These features have been specifically engineered to optimize unsupervised clustering of temporal sequences of viral load data that are asynchronous and noisy. Our findings demonstrate their success in identifying five types of viral load patterns often mentioned in the literature \cite{SupBenefits1,greub2002intermittent,rose2015comparison,main,de2006fatal,ylitalo2000consistent}. 

We have also proposed the novel centroid algorithm for giving a meaningful summarization to clustering results. Since the results of the algorithm are concise, it allows investigators to reconstruct the model to the programming language of their choice. Hence these results can positively aid HIV population case studies by giving precise definitions of the varying temporal viral load patterns, leading to an improvement in patient care.  This method would facilitate cross-comparison of research studies by providing a standard for viral load pattern classification.  Such standardization would be immensely useful in meta-analyses of diverse research reports \cite{etter2013recommendations, olsen2014risk,blaser2014impact,boender2015long,boerma2016suboptimal}.  It is possible that, in the future additional viral load patterns may emerge with, for example, the emergence of new HIV variants that are resistant to, or escape suppression of, current therapies.  The method is flexible enough to recognize different patterns, and thus categories of viral load responses.  Furthermore, this method is general enough that models could be trained on other viral infections that have patterns of natural or treatment related patient responses (e.g. hepatitis B and C, parvovirus B19).  Such applications would likely require defining new viral load time-series features that capture disease specific features.

A common practice in data analytics is to calculate the centroid as the average of the points \cite{abdi2009centroids,han2011data}, however we showed that the mean is not necessarily the best centroid. An advantage of the centroid algorithm is that we can choose the best centroid corresponding to the shape of the data, which may call for statistical tests to determine which shape best fits the data. Another advantage is that we can mathematically determine the amount of overlap between $n$-dimensional cluster spheres (i.e. viral load categories). With this property we can take different cohort studies, perform hierarchical clustering with centroid summarization on the new set of data, and then compare HIV viral load patterns (i.e. cluster comparison). Such comparisons may reveal influences of different patient care strategies or the relationship between different populations. Furthermore, since the centroid algorithm represents the data into centers and radii (bytes of data), another advantage of this algorithm is data reduction.

We have shown that, at the expense of model interpretability, centroid prediction may be improved with projected hyperplane normalization. The interpretability versus predictability problem is well known and often finds itself in deep learning papers  \cite{bologna1998symbolic,bologna2003model,intrator2001interpreting}. Interpretability is a desirable attribute in clinical classification systems, as it allows clinicians to integrate causal physiology and diagnostic information with data features in a way that may promote clearer bedside clinical reasoning.  The general understanding is that research vital to human safety must have an interpretable science, otherwise a ``black-box'' approach may be justified. Using an interpretable model for assigning viral load pattern membership may be advantageous for when a clinician wishes to use the assigned pattern membership to aid in making a critical clinical decision (e.g. choosing between treatment options). A convoluted model outcome may make such a decision more difficult \cite{Shickel_2017}.

Several caveats apply to our work.  As noted, this is a single center study, and thus our method should be tested with a much larger data set to cross-validate the categories represented by the clusters. In addition, our feature vector was designed specifically to suit the literature rather than objectively clustering the data using a time-series based clustering method \cite{Klapper_Rybicka_2001,bahadori2015functional,Kontaki_2008}. Also, some of our features are slightly collinear - with the greatest correlation coefficient being between IQR and wRR (-0.717). However, while HVLS and RVL both have a varied range of IQR, it is clear that the HVLS has greater wRR than RVL due to HVLS patients having a long consistent tail. Furthermore IQR helps distinguish the HVLS and the SLVL or SHVL class, hence both IQR and wRR are necessary despite the slight correlation. As another caveat to our work, we had to normalize time into \textit{number of days since admission} - meaning we lose the ability to look for seasonal or yearly patterns in the data. 

While we originally hypothesized the existence of six distinct viral load patterns, we found that the \textit{emergent VL} group may not be a pattern found in our data. Perhaps this is a consequence of a high rate of local patient engagement in therapy in this cohort study, or the era of heart therapy, or effectiveness of medication regiments and their access.  In different populations these conditions may not always exist (e.g. in areas where HAART is expensive and people may lose the ability to pay for it), for which the EVL pattern may indeed be significant.  Based on the formulation of the adj MD and wRR features, we hypothesize that a consequence of the grounding function is that any EVL pattern (if exists) will be grouped under RVL. This grouping may be appropriate as one can argue that the act of going from a suppressed state to a high VL state is a form of rebounding. 

Unsupervised clustering algorithms all have tuning parameters: the $k$ in k-Means or k-Medoids \cite{han2011data}, the radius (neighborhood size) in DBSCAN \cite{han2011data}, the density threshold and grid size in CLIQUE \cite{han2011data}, the branching factor in BIRCH \cite{han2011data}, or the $k$, $\alpha$, and minsize in CHAMELEON (perhaps the least sensitive of them) \cite{han2011data,karypis1999chameleon}.  Clustering results may change depending on the parameter chosen, revealing finer between-cluster differences as the number of clusters increase. The hierarchical clustering algorithm is no exception, but has the advantage that a proper cut-off can be easily visualized. Our method chose to make groups at a high level, but identification of important sub-clusters with a choice of a lower cut-off is also possible. For example, choosing a lower cut-off may reveal that the suppression group splits itself into categories with different rates of HIV viral load suppression during treatment. Researchers wishing to engineer a new feature vector for VL pattern segregation may find useful the supplementary material on features we considered but subsequently removed due to poor performance.

\section*{Conclusion}

We have proposed a set of four unambiguous features which have been successfully used in segregating five different types of viral load patterns: durably suppressed viral load (DSVL), sustained low viral load (SLVL), sustained high viral load (SHVL), high viral load suppression (HVLS), and rebounding viral load (RVL). We have also proposed a novel centroid-based clustering algorithm.  Applied to our data set, these methods resulted in clusters of temporal viral load patterns that map to clinically relevant treatment responses. The use of this algorithm may improve meta-analyses or population studies of viral load patterns by standardizing the classification of HIV patient categories. Furthermore, the segregation process used in this paper (i.e. discovering a set of domain specific features, performing unsupervised clustering, interpreting the results with a cluster summary) can be used to model other viral infections and the response of viral load levels over time to treatment or natural disease progression.

\section*{Supporting information}
\beginsupplement
\paragraph*{Fig S1}
\label{sfig:VLD}
{\bf Viral load distribution.} For each pair of viral load measurements, we calculate the change in days and the change in viral load counts for all patients and plot it as a scatter. The horizontal line of dots which appears between $0$ and $2$ are an artifact of using 20 and 48 in data to replace the ``Pos $<$20" and ``Pos $<$48" values which appeared in our data. The sequential range of viral load measurements shows that VL measurements taken within 10 days of each other may vary by $\pm 10^5$ copies/mL.

\paragraph*{Fig S2}
\label{sfig:outline}
{\bf Paper Outline.} The light green box indicates the training step and the light red box indicates the testing step. (A) Feature extraction (B) Normalization of extracted features. Each feature has a unique linear transformation for normalization. (C) Unsupervised hierarchical clustering into clouds (the six different VL patterns: SHVL, SLVL, DSVL, HSVL, Emergence, Slow Suppressors) (D) Centroid summarization taking as input the cluster labels from hierarchical clustering and normalized data (while storing $f$ [the normalization functions]) (E) Partition viral load data into retained proportions. (F) Use only the amount of data we have real data points for. (G) Normalize the features into the same transformed space the training was performed on. (I) Perform \textit{radial normalization} classification on the normalized features to classify the partially retained VL pattern.

\paragraph*{Fig S3}
\label{sfig:bivariate}
{\bf Patient feature extraction.} Feature extraction on 1576 patients displayed as 2D splicing of the 4 dimensional feature space. Each splice plots a dimension versus another in the form of a scatter plot.

\paragraph*{Fig S4}
\label{sfig:DecisionTree}
{\bf Decision Tree.} While some useful rules may be pruned, the tree is otherwise complicated and difficult to draw useful conclusions from.

\paragraph*{Fig S5}
\label{sfig:DTCS}
{\bf Decision Tree Classification Stability.} Classification stability results using decision tree learned on 100\% retained data. The title of each plot corresponds to which category of patients are being tested. The dashed white line is the 80\% membership assignment probability line, included for reference. The solid white line represents the number of patients included in the prediction at the point of retained viral load data (corresponding to the right axis) which changes as a result of the $r \geq 3$ restriction.

\paragraph*{Fig S6}
\label{sfig:SVMCS}
{\bf SVM Classification Stability.} Classification stability results using SVM learned on 100\% retained data. The title of each plot corresponds to which category of patients are being tested. The dashed white line is the 80\% membership assignment probability line, included for reference. The solid white line represents the number of patients included in the prediction at the point of retained viral load data (corresponding to the right axis) which changes as a result of the $r \geq 3$ restriction.

\paragraph*{Fig S7}
\label{sfig:kNNCS}
{\bf $k$-Nearest Neighbor Classification Stability.} Classification stability results using $k$-Nearest Neighbor ($k = 5$) learned on 100\% retained data. The title of each plot corresponds to which category of patients are being tested. The dashed white line is the 80\% membership assignment probability line, included for reference. The solid white line represents the number of patients included in the prediction at the point of retained viral load data (corresponding to the right axis) which changes as a result of the $r \geq 3$ restriction.

\paragraph*{Fig S8}
\label{sfig:PCShn}
{\bf Polyhedral CMuPHN Classification Stability.} Classification stability results using the polyhedral method for detecting center and projected hyperplane normalization method for classification. The title of each plot corresponds to which category of patients are being tested. The dashed white line is the 80\% membership assignment probability line, included for reference. The solid white line represents the number of patients included in the prediction at the point of retained viral load data (corresponding to the right axis) which changes as a result of the $r \geq 3$ restriction.

\paragraph*{Fig S9}
\label{sfig:centmethods}
{\bf Centroid Methods.} Gives a visual of how the seven methods work on an example point set. The green target signifies the exact center which is found according to the different methods in our algorithm.

\paragraph*{Data S1}
\label{sdata:VLdataset}
{\bf Viral load data.} The data set used for this study is provided in a completely deidentified format. The data is in a csv format where the first column represents a unique subject, with a random identifier. The subsequent values are as $t_{i,j}, VL_{i,j}$, where $t_{i,j}$ is the time from a universal $T_0$ for the VL measurement $j$ for patient $i$, and $VL_{i,j}$ is the corresponding VL measurement.  Each record (row) is of a unique length, depending on the number of VL measurements present for that subject.  The code used to analyze the data can be found on github at https://github.com/SamirRCHI/Viral\_Load\_Data\_Categorization.

\newpage

\begin{table}[htbp]
  \caption{\textbf{Correlation Coefficient Matrix of Features}}
    \begin{tabular}{l|rrrr}
    \toprule
          & \multicolumn{1}{l}{Area} & \multicolumn{1}{l}{wRR} & \multicolumn{1}{l}{Adj MD} & \multicolumn{1}{l}{IQR} \\
          \midrule
    Area  & 1     & -0.5639 & -0.0155 & 0.5243 \\
    wRR   & -0.5639 & 1     & 0.0325 & -0.717 \\
    Adj MD & -0.0155 & 0.0325 & 1     & -0.2379 \\
    IQR   & 0.5243 & -0.717 & -0.2379 & 1 \\
    \bottomrule
    \end{tabular}%
  \label{stab:correlationcoefficient}%
\end{table}%

\vspace{5mm}

\begin{table}[htbp]
  \caption{\textbf{Performance Comparison with Polyhedral CMuRN}}
    \begin{tabular}{l|llll}
    \toprule
          & \textbf{Decision Tree} &  \textbf{SVM}   &  \textbf{k-NN,k=5} &  \textbf{PH CMuPHN} \\
          \midrule
    \textcolor{HVLS}{HVLS}  & (0.0, 0.016) & (0.0367, 0.0496) & (0.0224, 0.0426) & (0.0171, 0.0287) \\
    \textcolor{SHVL}{SHVL}  & (-0.0263, 0.04) & (-0.0263, 0.0385) & (0.0, 0.0769) & (-0.1818, -0.1154) \\
    \textcolor{DSVL}{DSVL}  & (0.0038, 0.0143) & (0.0, 0.0096) & (0.0028, 0.012) & (-0.0024, 0.0) \\
    \textcolor{RVL}{RVL}   & (-0.069, -0.04) & (-0.0657, -0.0438) & (-0.0576, -0.0419) & (0.0, 0.0181) \\
    \textcolor{SLVL}{SLVL}  & (0.0068, 0.0394) & (0.0293, 0.0484) & (0.0445, 0.0633) & (0.0352, 0.0542) \\
\bottomrule
    \end{tabular}%
    \\\\
    PH = Polyhedral\\
    The results are the Quartiles ($Q_1$,$Q_3$) of the difference between the stated algorithm.\\
    A negative score means Polyhedral CMuRN performed better.\\
    Only $ri$ instances where Patients $\geq 20$ were used in this comparison.\\
  \label{stab:quartiles}%
\end{table}%

\newpage
\section*{Acknowledgments}
This work was partially funded by the University of Rochester Clinical and Translational Science Institute, supported in part by grants UL1 TR002001, and TL1 TR002000 from the National Center for Advancing Translational Sciences (NCATS), a component of the National Institutes of Health (NIH).  This publication was also made possible through core services and support from the University of Rochester Center for AIDS Research (CFAR), an NIH-funded program (P30 AI078498).  The content is solely the responsibility of the authors and does not necessarily represent the official views of the National Institutes of Health (NIH).

We would also like to thank Yusuf Bilgic (State University of New York at Geneseo), for discussions regarding the statistical analyses.


%
%
%

\newpage

\newpage
\section*{Supplementary Material Section}

\subsection*{Review of Existing Viral Load Categorization Methods}

\subsubsection*{Greub et. al. LLVR}

Greub et. al. were particularly focused on detecting low level viral rebound (LLVR) in patients \cite{greub2002intermittent}. The following procedure was used to categorize the patients of their study:

If the patient has two consecutive viral load measurements (VLM) less than 50, within a 24 week period, and they have two VLM after this consecutive pair occurs, then the viral load data for that patient is considered for further analysis. For the patients that meet this criteria, their viral load measurements following the consecutive pair are viewed: if their maximum VLM is greater than 500, then they are categorized as `Viral Failure', if they are between 51-500 then they are categorized as `LLVR', otherwise they are labeled as `DSVL'. Greub et. al. left out the patients which did not meet the consecutive pair criteria from their categorization, hence in our comparative study we will group them under `Unspecified'.

\subsubsection*{Rose et. al. SMVL/RMVL}

The focus of Rose et. al. was to investigate the use of several frameworks in categorizing suppressed versus not-suppressed viral load \cite{rose2015comparison}. First they omitted the patients from their study whom were virally suppressed at baseline, where they define viral suppression as $< 200$ copies/mL because they were found to have no substantial variation in their viral loads. In our comparative analysis we label them as `Baseline $< 200$'. Then, from the remaining patients they categorize them as either achieving suppression or not-suppressed using an 8 month window centered around month 24 after start of VLM (18-30 months). They considered five different frameworks, which we describe below:

\begin{enumerate}
\item[] \textbf{SMVL omit-participant}: If the closest VLM to month 24 is $< 200$ then the patient is labeled as `Suppressed', otherwise `Not Suppressed'. However if this closest VLM is outside the range of 18-30 months, then they are labeled as `Ommitted'.
\item[] \textbf{SMVL set-to-failure}: Similar to omit-participant, however if the the closest VLM is outside the range of 18-30 months, then they are labeled as `Not Suppressed'.
\item[] \textbf{SMVL closest-VL}: The patient is labeled according to their closest VLM to month 24 regardless of whether the VLM is contained in the window.
\item[] \textbf{RMVL repeat binary}: We do not use this method in our comparative analysis because its purpose is to classify each individual VLM as suppressed or not suppressed (rather than the patient), which is not the goal of this paper.
\item[] \textbf{RMVL repeat continuous}: The $\log_{10}$ of the VLMs are modeled as a continuous linear function with its intercept fixed at the baseline viral load. In our implementation we add 1 to each VLM to avoid $\log_{10}(0)$. Then the patient is categorized as `Suppressed' if the model predicts the patient to have a viral load $< 200$ copies/mL at month 24, otherwise they are labeled as `Not Suppressed'.
\end{enumerate}

\subsubsection*{Terzian et. al. SHVL}

The objective of Terzian et. al. was to develop a method of categorizing a patient as DSVL or SHVL for the purpose of monitoring successful ART uptake \cite{main}. Their procedure for categorizing patients is as follows:

If the maximum viral load of the patient is $\leq 400$ copies/mL then the patient is labeled as `DSVL'. If the patient instead has two consecutive viral load measurements $\geq$100,000 copies/mL, then the patient is labeled as `SHVL'. For our analysis all other patients are labeled as `Unspecified'.

\subsubsection*{Phillips et. al. Viral Rebound}

The aim of Phillips et. al. was to characterize virological response to ART \cite{phillips2001hiv}. While the statistical methods proposed by Phillips et. al. went beyond categorizing patients, they composed a method to identify two populations of HIV patients (Viral Failure and Viral Rebound):

Only patients who have at least one VLM within the range 24-40 weeks are included in the categorization, all other patients are labeled as `Omitted' (similar to SMVL omit-participant). Phillips et. al. chose the 24-40 week range for a different part of their statistical analysis; however, in our categorical implementation, we modify the range to 24-32 weeks to build coherent categories according to the procedure they outline (as we are about to describe). Phillips et. al. choose to use 32 weeks as the point of observing viral load levels because they argue viral load is expected to decline to 500 copies/mL by week 32 if it is going to do so \cite{phillips2001hiv}. Thus patients who never achieve VL $<$ 500 copies/mL within 32 weeks are labeled as `Viral Failure'. If the patient does achieve VL $<$ 500 at one point within 32 weeks, but VLM closest to 32 weeks is $>=$ 500 copies/mL, then this patient is omitted from the categorization. If, however, the patient's closest VLM to 32 weeks is $<$ 500 copies/mL, then if the patient has two consecutive VLM $>=$ 500 copies/mL within 32 weeks, then the patient is labeled as `Viral Rebound'. Phillips et. al. do not describe them further the patient who did not meet this consecutive pair criteria, however for our implementation we will label them as `Suppressed'.\subsection*{Considered but Removed Features}
There were several features which were thought to have significance in segregating viral load patterns but did not make it into our feature vector. We had to be careful of extracting features which may be collinear as it would cause a shift in the weighting of features. These collinear features are too many to list here. We mention several excluded features which were generally unreliable or created overlap between classes that should be unrelated:

\begin{enumerate}
\item \textit{Minimum.}

\item \textit{Maximum.}

\item \textit{Baseline VL.} 

\item \textit{Last VL.}

\item \textit{Rate of Change.} We tried several ways to calculate this feature: 1) Mean of the first derivate of $\vv{VL}_p$ with respect to $\vv{t}_p$, 2) Median of the first derivative, 3) Rate of change between first and last measurements, 4) Fitting a piecewise regression with one knot and averaging the results of the slope- with the idea that HVLS will have a clear elbow.

\item \textit{Correlation Coefficient.} From the fitted piecewise regression (one knot) we also tried averaging the results of the correlation coefficient, again with the idea that HVLS will have a clear elbow.

\item \textit{Change in Concavity.} Calculated as sum of the absolute changes in concavity with the assumption of equally spaced time (to adjust for issues found in calculating rate of change):

\[
\vv{dVL}_i = \vv{VL}_{i+1} - \vv{VL}_i
\]

\[
\vv{concavity}_i = \vv{dVL}_{i+1} - \vv{dVL}_i
\]

\[
Change = ||\vv{concavity}||_1
\]

\item \textit{Positive Difference.} We calculate this as the sum of all $\vv{dVL}_i \geq 0$ normalized by the number of elements satisfying the condition.

\item \textit{Negative Difference.} We calculated this as the sum of all $\vv{dVL}_i < 0$ normalized by the number of elements satisfying the condition.

\end{enumerate}

\subsection*{Centroid Detection Methodologies}
Centroid detection is a problem which several machine learning algorithms attempt to solve, such as Support Vector Machine (SVM), Bayesian Point Machines (BPM), Analytic Centre Machines, k-Means, BIRCH, among others \cite{maire2003algorithm, han2011data}. The center of a cloud of samples is generally considered the average \cite{abdi2009centroids}, but is still a matter of interpretation. We attempt to find cluster centers in seven different ways (\nameref{sfig:centmethods}):

\begin{enumerate}
\item \textit{Mean.} Calculated as the average of each dimension (feature).

\item \textit{Median.} Calculated as the median of each dimension (feature).

\item \textit{Best Representative.} The best representative is taken to be the sample whose maximum distance to all the other samples is a minimum.

\item \textit{Bounding Box.} If the cloud of points naturally form an $n$-dimensional box, then calculating the middle of the minimum and maximum of each dimension would lend itself to be the center representation of the cloud.

\item \textit{Smallest Disk.} If the cloud of points instead naturally form an $n$-dimensional sphere, then it makes most sense to solve the smallest enclosing disk problem. There are many algorithms that have been proposed to solve this problem \cite{welzl1991smallest,fischer2003fast,nielsen2004approximating}, where we found the best solution in our implementation of the algorithms to be that of Fischer's fast smallest-enclosing ball, where he utilizes the properties of the hull and affine spaces to walk the center to its optimum location \cite{fischer2003fast}. They proposed to initiate the center by choosing a random sample, however this caused the algorithm to yield differing solutions on each run. Through experimentation we found that when we initialized the algorithm using the sample found from best representative method, the algorithm consistently converged at the solution with the smallest radii, compared to using any other sample as the initial center. Hence in our formulation of Fischer's algorithm we deploy the best representative method as a subroutine for initiation. 

\item \textit{Polyhedral.} The trouble with the bounding box and smallest disk approximations for the center is that they are designed to work well under circumstances where the clusters naturally form either a $n$-dimensional box or sphere respectively. We can relax this constraint by removing this assumption. Rather we propose to find the convex hull of the data points and then find the centroid of the hull, in other words, we find a $n$-dimensional polyhedron to fit the cloud. Then, given the intersection of the finite half-spaces of the convex hull, we can calculate the exact centroid of the polyhedron- where we use a slight modification of Maire's algorithm \cite{maire2003algorithm}. The hyperplane equations are calculated in Python using the \mcall{ConvexHull} and \mcall{Delaunay} packages within \mcall{SciPy}.

\item \textit{Push and Pull.} The six methods mentioned are not designed to maximize the distance between each cluster center while minimizing the distance within the cluster. In other words, the above six methods calculate the center of each point cloud without any dependencies on the other clouds. We propose a push and pull method, inspired by force-directed graph drawing using Fruchterman-Reingold's algorithm \cite{kobourov2012spring} to include dependencies on the center of each cloud to improve performance on centroid learning. That is, let every sample within the cluster have a pulling force on the center following Hooke's Law, $F = kd$, and let every sample outside the cluster have a pushing force on the center following a modified version of Coulomb's Law, $F = k_e\frac{q_1q_2}{d^2}$. In our case, $d$ represents distance between the center and sample, and the constants $k, k_e, q_1,$ and $q_2,$ equal to 1 in order to yield equally weighted pulling and pushing. For each cluster, we initialize the center as the mean of the cluster followed by iteratively finding the center until achieving a dynamic equilibrium. We initiate at the mean of the cluster samples because the spring-like pulling of each sample on the center is the same as the mean of the samples pulling on the center with a spring constant equal to the number of samples (proved below).\\
\hspace{1.5mm}

\textbf{Proposition} \textit{The combined spring-like pull of each sample, $S_i$, on a center, C, with a spring constant of 1, is same as a spring-like pull of the mean of the samples, $\overline{S}$, on C with a spring constant equal to the number of samples pulling on C.}\\
\hspace{1.5mm}

\textit{Proof.} Let the sample $S_i$ and the center $C$ be $n$-dimensional vectors such that $S_i$ has a spring-like pulling force on $C$ with a spring constant of 1, that is, $F_i = \frac{||S_i-C||}{||S_i-C||}(S_i - C) = S_i - C$, where $\frac{1}{||S_i-C||}(S_i-C)$ is the direction of the force and $||S_i-C||$ is the magnitude, or distance between $S_i$ and $C$. Notice the mean of the samples $\overline{S} = \frac{\sum_{i=1}^n S_i}{n}$, where $n$ is the total number of samples, 
then observe that the sum of the spring-like forces acting on the center is given by
\[
\begin{split}
F & = \sum_{i=1}^n F_i \\
  & = \sum_{i=1}^n (S_i - C)\\
  & = \sum_{i=1}^n S_i - \sum_{i=1}^n C \\
  & = (\sum_{i=1}^n S_i) - nC \\
  & = n(\frac{\sum_{i=1}^n S_i}{n} - C) \\
  & = n(\overline{S} - C)
\end{split}
\]
This yields we have that the sum of all the spring-like forces on $C$ is the same as the mean average of the samples having a spring-like pull on $C$ with a spring constant of $n$, which is the number of samples pulling on C. \qedsymbol \\
\hspace{1.5mm}

Observe that some clusters have a low number of samples pulling at the center relative to all the samples having a pushing effect. Hence, if $n$ is small, then we will find that the center is pushed more than desired, and if $n$ is large, then the pull towards the average will be too strong and hence the center will be no different from the average. We propose to replace $n$ with the number of samples having a pushing effect on the center.
\end{enumerate}

\end{document}